\documentclass[preprint,showpacs,preprintnumbers,amsmath,amssymb]{revtex4}
\usepackage{epsfig}
\begin{document}

\title{Reply to Comment on ``Quantum Coherence between High Spin Superposition States of Single Molecule Magnet Ni$_4$"}
\author{E. del Barco and A. D. Kent}
\affiliation{Department of Physics, New York University, 4
Washington Place, New York, NY 10003}
\begin{abstract}
Here we respond briefly to a comment on our work in
arXiv:cond-mat/0405501.
\end{abstract}
\date{\today}

\maketitle

A recent Comment \cite{WernsdorferComment} was posted on cond-mat
on our paper entitled ``Quantum Coherence between High Spin
Superposition States of Single Molecule Magnet Ni$_4$"
\cite{delBarco}. The main points of this comment can be summarized
as follows:

1) That our paper reported experimental studies on the $S=4$
single molecule magnet Ni$_4$ using a method that combines high
sensitivity magnetic measurements with microwave spectroscopy,
which was similar to the work posted on cond-mat a few weeks
earlier on the $S=1/2$ molecule  V$_{15}$ \cite{WernsdorferCM}.
The implication was that this early posting gives Wernsdorfer
claim to being the `originator' of this magnetic measurement
method.

2) That pulsed-radiation relaxation experiments, like those we
present, do not give directly the spin-lattice relaxation time
T$_1$ but the spin-phonon-bottleneck time. As a consequence our
finding of an increase of T$_1$ with frequency is not contrary to
general ideas.

3) The final claim is that ``... all statements concerning the
observation of quantum coherence'' have not yet demonstrated.

We address these points in the order above.

First, we started planning and discussing these experiments three
years ago, including ordering the necessary equipment \cite{note}.
The experiments on $Ni_4$ were conducted over the last six months.
The cited work was presented in an invited talk
\cite{delBarcoTALK} at the APS March Meeting and at an earlier
conference \cite{KentTalk}. Our paper was submitted to Science
after the March Meeting (on April 9$^{th}$ 2004, one week before
ref. \cite{WernsdorferCM}) and shortly thereafter to Physical
Review Letters. We also note that while our paper was under
consideration for publication, an interesting related work by Bal
et al. \cite{Friedman} appeared. The timing of these publications
and talks makes it clear that these papers correspond to
independent research carried out in {\it parallel}. These
experiments and other earlier publications reflect the continued
and growing interest of the community in coherent quantum
phenomena in SMMs \cite{delBarco2,Hill}. The ideas of such
combined magnetic/microwave experiments were actively discussed by
many in the SMM community
\cite{note,Tejada,WernsdorferPolarization}, and similar
experiments were published some time ago, which investigated
quantum superposition states in superconducting quantum
interference devices \cite{Mooij,Friedman2}.

In our experiments high sensitivity Hall magnetometry was combined
with microwave spectroscopy. We note that Hall magnetometers have
recently been demonstrated to have sensitivities of $5 \times
10^{5}$ spins \cite{vonMolnar} and higher spin sensitivities are
certainly possible, with material and device improvements. There
is also no fundamental reason that Hall magnetometers cannot be
used in ultra-fast time resolved measurements ($\sim$ 10 GHz),
after all, the same semiconductor heterostructure materials used
for Hall-magnetometers are used in high-speed circuitry. We note
that time resolved magnetic measurements at the nanosecond time
scale with micro-SQUIDs have not yet been demonstrated, only fast
measurements with SQUIDs specialized for another purpose
\cite{Chiorescu_SQUIDs}. The main advantage of Hall magnetometry
is the possibility of high sensitivity magnetic measurements over
a the wide range of temperatures and applied fields (including,
very large applied fields).
\begin{figure}
\begin{center}\includegraphics[width=13cm]{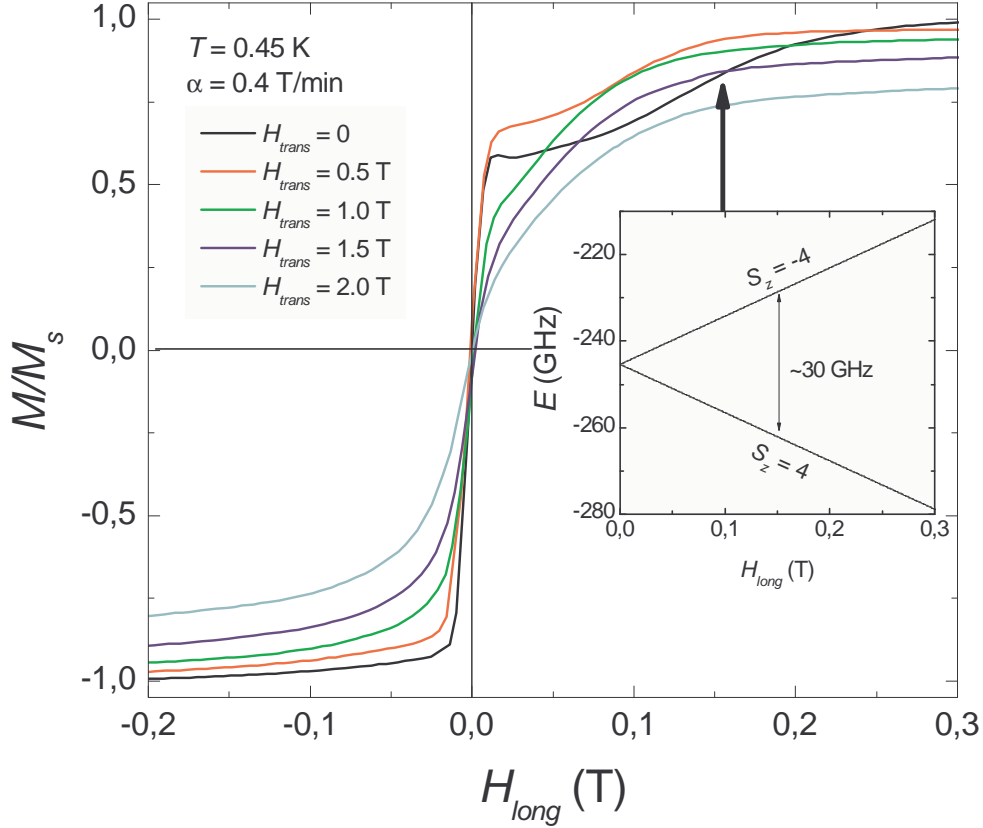}
\caption{Magnetization curves of Ni$_4$ obtained sweeping at a
constant rate, $\alpha=0.4 $ T/min, a longitudinal magnetic field
in the presence of different transverse fields. The inset shows
the energy of the lowest levels as a function of the longitudinal
field.}
\end{center}
\end{figure}

As alluded to in the Comment, spin-phonon relaxation processes are
poorly understood in SMMs (see, for example, ref.
\cite{Chudnovsky}). It is certainly true that a
spin-phonon-bottleneck can affect the relaxation time in SMMs
\cite{Chiorescu_Bottleneck} . We considered this possibility. The
small phonon heat capacity and small coupling of sample phonons to
the (cryostat) environment can create an out-of-equilibrium
distribution of phonons \cite{Abragam}. We have observed this
phenomenon in our Ni$_4$ experiments (not reported in ref.
\cite{delBarco}) as a plateau in the magnetization hysteresis
curve (magnetization lower than its saturation value) after
sweeping thought zero field (resonance $k=0$) in the absence of
transverse fields (see fig. 1) \cite{Wernsdorfer_unplugged}. The
magnetization approaches its equilibrium value when the
longitudinal field is $\sim0.15$ T. At this field the separation
between lowest levels is about 30 GHz (see inset in fig. 1). At
this frequency, the density of available phonon modes is
apparently sufficient to allow energy relaxation on the time scale
of the hysteresis sweep. Importantly, this plateau gradually
disappears with increasing magnitude of an applied transverse
field. This is due to the fact that a transverse field increases
the energy splitting at the resonance, allowing the system to
reach a state in which more and more phonons become available (the
distribution of thermal phonons reaches a maximum at
$3k_BT/h\sim20-40$ GHz).

It is clear from this line of reasoning that since the density of
phonon modes increases with frequency ($\sim\omega^2$), the
effective relaxation time for a process limited by the
phonon-bottleneck should {\it decrease} with increasing radiation
frequency ($\tau \sim 1/\omega^2 $). However, in our experiments
with pulsed microwave fields we observe the opposite behavior (see
Fig. 3C of Ref. \cite{delBarco}), which is not understood.
Therefore, and as claimed in Ref. \cite{delBarco}, the observed
dependence of the relaxation time on frequency is opposite that
expected based on the increasing phase space available for phonon
generation.

On the last point, we note that our experiments with microwave
radiation in Ni$_4$ have been carried out in the presence of
transverse fields. For this reason, our experiments
\cite{delBarco} differ significantly from those discussed above
\cite{WernsdorferCM,Friedman}. A magnetic field applied
perpendicular to the uniaxial anisotropy axis of the molecules
breaks the degeneracy between symmetric and antisymmetric
superpositions of $S_z$ spin projections, producing an energy
splitting between the states, known as the tunnel splitting. We
note that in Refs. \cite{WernsdorferCM,Friedman} the energy
splitting between states is linear with applied field and simply
corresponds to the Zeeman splittings between $S_z$ states. The
tunnel splittings are not directly resolved in these experiments.
In our experiments these splittings are directly resolved and
reflect the formation of high spin-superposition states. The lower
bound on the coherence time has been determined for these states
to be $ \sim 1$
 ns, an order of magnitude higher than the time scale set
 by radiation frequency used in the experiment.  The observation of quantum
coherence between high spin-states clearly motivates future work
in our group and others that will examine the quantum dynamics in
the frequency and time domain, such as Rabi experiments.

{\bf Acknowledgements:} We thank Dr. Wernsdorfer for making his
comments known to us before they appeared on cond-mat. We
acknowledge our collaborators, E. C. Yang and D. N. Hendrickson
and co-authors of ref. \cite{delBarco} for their important
contributions to this research program. This research was
supported by the NSF (Grant Nos. DMR-0103290, 0114142, and
0315609).

\end{document}